\documentclass{article}
\usepackage{amsmath,amssymb,amsfonts}
\usepackage{algorithmic}
\usepackage{graphicx}
\usepackage{algorithm,algorithmic}
\usepackage{hyperref}
\hypersetup{hidelinks=true}
\usepackage{textcomp}
\usepackage{biblatex}
\def\BibTeX{{\rm B\kern-.05em{\sc i\kern-.025em b}\kern-.08em
    T\kern-.1667em\lower.7ex\hbox{E}\kern-.125emX}}
\begin{document}
\title{Efficient Quality Control of Whole Slide Pathology Images with Human-in-the-loop Training}
\author{Abhijeet Patil, Harsh Diwakar, Jay Sawant, Nikhil Cherian Kurian, \\ Subhash Yadav, Swapnil Rane, Tripti Bameta, Amit Sethi }

\maketitle

\begin{abstract}

Histopathology whole slide images (WSIs) are being widely used to develop deep learning-based diagnostic solutions, especially for precision oncology. Most of these diagnostic softwares are vulnerable to biases and impurities in the training and test data which can lead to inaccurate diagnoses. For instance, WSIs contain multiple types of tissue regions, at least some of which might not be relevant to the diagnosis. We introduce HistoROI, a robust yet lightweight deep learning-based classifier to segregate WSI into six broad tissue regions -- epithelium, stroma, lymphocytes, adipose, artifacts, and miscellaneous. HistoROI is trained using a novel human-in-the-loop and active learning paradigm that ensures variations in training data for labeling-efficient generalization. HistoROI consistently performs well across multiple organs, despite being trained on only a single dataset, demonstrating strong generalization. Further, we have examined the utility of HistoROI in improving the performance of downstream deep learning-based tasks using the CAMELYON breast cancer lymph node and TCGA lung cancer datasets. For the former dataset, the area under the receiver operating characteristic curve (AUC) for metastasis versus normal tissue of a neural network trained using weakly supervised learning increased from 0.88 to 0.92 by filtering the data using HistoROI. Similarly, the AUC increased from 0.88 to 0.93 for the classification between adenocarcinoma and squamous cell carcinoma on the lung cancer dataset. We also found that the performance of the HistoROI improves upon HistoQC for artifact detection on a test dataset of 93 annotated WSIs. The limitations of the proposed model are analyzed, and potential extensions are also discussed. Code available at \href{https://github.com/abhijeetptl5/historoi}{https://github.com/abhijeetptl5/historoi}.

\end{abstract}

\section{Introduction}
\label{sec:introduction}

Computational pathology -- the use of computational (specifically, deep learning) techniques for diagnostic analysis of digital pathology whole slide images (WSIs) -- is well on its way to being incorporated for clinical diagnosis in the coming few years. However, variance in the quality of tissue and slide preparation as well as scanning of WSIs, remains a concern in ensuring that deep learning models trained on curated datasets during the development phase work well during deployment as well. Due to this reason and also due to the lack of sufficient computational power and software libraries that can handle WSIs, most of the research on computational pathology in the last few years was concentrated on the analysis of clean and carefully curated datasets of patches (sub-images of manageable size) extracted from histopathology WSIs~\cite{breakhis, bach}. With increasing computational resources, research on using WSIs for diagnosis has gained traction, where either region-level annotations or weakly supervised learning methods are used. The former is laborious and prone to human bias and errors due to the large (gigapixel) size of WSIs. On the other hand, while weakly supervised learning can eliminate the burden of regional annotations and reduce human bias~\cite{clam,selfsupwsi,jpath_deepak}, it can introduce its own biases as we lose control over the accidental association of slide-level labels to even some of the irrelevant patches in a WSI.

In order to improve patch selection for various levels of supervision, we propose a fast, semi-automated, and iterative method that combines dataset preparation and model training to classify WSI patches into various commonly used tissue segment classes. Our method has the following advantages. Firstly, it keeps a human-in-the-loop for cluster-level weakly supervised annotations. Secondly, its number of iterative model refinements can be manually controlled to trade-off between the quality of its results and annotation effort. Thirdly, the trained model generalizes from only one dataset to unseen datasets and organs for patch-level classification. And lastly, we show that the results of weakly supervised learning algorithms can be improved using the patch classification based on our model as a quality-control (QC or filtering) strategy for multiple problems and organs.

\begin{figure}[h]
	\centering
		\includegraphics[width=0.9\textwidth]{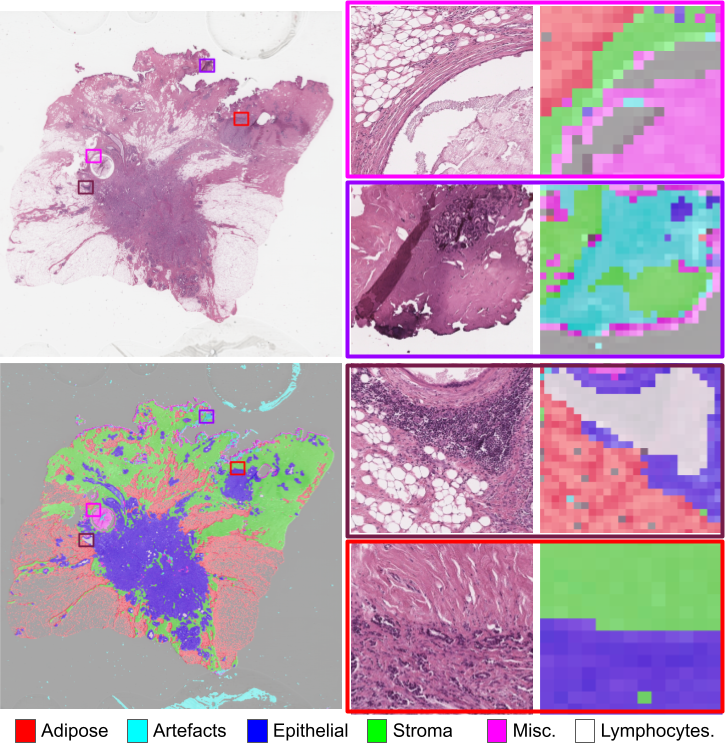}
	\caption{An example of segmentation results on a whole slide image using HistoROI into epithelium, stroma, lymphocytes, adipose, artifacts, and miscellaneous areas}
	\label{input-output}
\end{figure}

We elaborate upon the problem that we have solved as follows. Many diagnostic tasks, such as identifying cancer in breast tissue or subtypes of renal carcinoma, depend on the structure of epithelial cells~\cite{bright, clam}. Structural features of the stroma have been shown to be related to the survival and recurrence of breast cancer~\cite{stroma, stroma2}. Similarly, the interaction between epithelial cells and lymphocytes is used as an indicator of survival in breast cancer patients~\cite{tiger}. However, most of these tasks do not depend on the analysis of the adipose (fat) regions or the artifacts. In fact, it has been shown in multiple studies that the performance of histopathology classification algorithms degrades if artifacts are not removed from the analysis~\cite{jbhi, stresstest,jco_nikhil}. Artifacts in a WSI can get introduced during various stages, including improper tissue fixation, irregularities in thickness while cutting tissues, tissue folds introduced while putting tissue sections on glass slides, over- or under-staining, the introduction of foreign objects, the introduction of air bubbles while putting slide cover, marking with a pen over the tissue area, and improper focus while scanning. Artifact removal can further improve both strongly and weakly supervised classification pipelines. While neural networks take patches as input instances, in weakly supervised learning, the labels are available only at the bag level, where \emph{bag} refers to a set of patches extracted from a WSI. Introducing too many irrelevant instances in a bag, such as those containing artifacts, can confuse neural networks.

Though a variety of histopathology datasets are publicly available on platforms, such as TCGA~\cite{tcga} and grand challenges~\cite{grandc}, there is no general and reliable solutions available for pre-processing patches pathology-specific labels, such as artifacts. A few solutions proposed earlier for the detection of artifacts in WSIs~\cite{histoqc, histoqc2, pathprofiler} provide good results only for specific organs and datasets. Detection of artifacts in WSI is challenging because the range of visual features associated with artifacts is very broad and prone to subjectivity. Another challenge in developing a model for whole slide image segmentation or patch classification is the unavailability of training data containing enough variations for generalization. 

In this study, we present a lightweight classifier -- HistoROI -- to segregate patches of WSIs into one of the following classes -- epithelium, stroma, lymphocytes, adipose, artifacts, and miscellaneous. Figure~\ref{input-output} shows an input-output map of HistoROI on a sample WSI. HistoROI is trained with a novel human-in-the-loop training paradigm to ensure variation in training data and robust classification and to reduce the annotation and labeling burden on human experts. Our method runs a few iterations of over-clustering of patches, label assignment to pure clusters, sampling from the impure clusters for labeling, and improving the classifier.

We have validated HistoROI on a patch-based colon cancer dataset~\cite{crc100k} to show its ability to generalize to organs other than the breast tissue on which it was trained. We have also demonstrated the applicability of the proposed model for detecting artifacts in WSIs. Further, we have shown improvement in the performance of the weakly-supervised learning models for the CAMELYON~\cite{camelyon} and TCGA-Lung~\cite{tcga} datasets after using HistoROI to remove obviously unwanted patches from WSI bags.

Specifically, our contributions are the following:
\begin{itemize}
    \item We introduce HistoROI, the classification model trained using the human-in-the-loop training paradigm. This method makes optimal use of the annotators' time by ensuring presenting them only difficult and diverse patches. [Publically available \href{https://github.com/abhijeetptl5/historoi}{HistoROI}.]
    \item We prepared a dataset of histopathology patches with six different classes. This dataset contains more than 2 million patches with enough variation for each class. It includes a variety of artifacts and can be used to prepare quality control (QC) solutions. [Publically available \href{https://github.com/abhijeetptl5/historoi/tree/main/training_data}{HistoROI-dataset}.]
    \item  We show the utility of the proposed model on the CAMELYON and TCGA-Lung datasets. We have used the proposed model as a pre-processor for a classification algorithm and showed an increase in AUC from 0.88 to 0.92 on the CAMELYON dataset. We also explored the effect of quality control (QC) on the automatic WSI diagnosis pipeline. We have conducted a similar study on the TCGA-Lung dataset and demonstrated promising results. Though a few studies have been done on the effects of QC on diagnosis using patches of WSIs using simulated data~\cite{jbhi, stresstest, jco_nikhil}, to the best of our knowledge, this is the first study to show results on WSIs using real artifacts. We also present improvement in the explainability of weakly supervised learning models with the help of HistoROI.
    \item To validate HistoROI for QC, we annotated the foreground tissue region for 93 WSIs with a binary mask. This dataset can also be useful for the development and validation of novel QC solutions for WSIs. [Publically available \href{https://github.com/abhijeetptl5/historoi/tree/main/tcga_tissue}{TCGA-4Org}.]
\end{itemize}

We introduce the datasets used for training and evaluation in Section \ref{sec_datasets}. We describe our human-in-the-loop active learning method in more detail in Section \ref{sec_methods}. We share the results of using HistoROI as a pre-processing tool, its applicability to the colon cancer dataset and using it for QC in Section \ref{sec_results}. We conclude in Section \ref{sec_conclusion}.

\section{Datasets} \label{sec_datasets}

\begin{table*}
\centering
\begin{tabular}{|c|l|c|}
\hline
\textbf{Dataset}                & \multicolumn{1}{c|}{\textbf{Composition}}                                                                                                                                                                 & \textbf{Purpose}                                                                                          \\ \hline
BRIGHT                          & \begin{tabular}[c]{@{}l@{}}2,169,355 Patches from 50 WSIs\\ 787,168 - Epithelium, 863,989 - Stroma\\ 98,293 - Lymphocytes,  245,525 - Adipose\\ 127,393 - Artifacts, 46,987 - Miscellaneous\end{tabular} & \begin{tabular}[c]{@{}c@{}}Training and validation of \\ HistoROI\end{tabular}                           \\ \hline
CRC-100k                        & \begin{tabular}[c]{@{}l@{}}100,000 patches of colon cancer\\ 10407 - ADI, 8763 - NORM, 10566 - BACK, \\ 11557 - LYM, 14317 - TUM, 13536 - MUS,\\ 8896 - MUC, 10446 - STR, 11512 - DEB\end{tabular}       & \begin{tabular}[c]{@{}c@{}}Validation of HistoROI on \\ external dataset\end{tabular}                    \\ \hline
TCGA-4Org                     & \begin{tabular}[c]{@{}l@{}}93 multi-organ WSIs annotated with \\ annotated tissue region\end{tabular}                                                                                                    & \begin{tabular}[c]{@{}c@{}}Application of HistoROI \\ for QC\end{tabular}                                \\ \hline
CAMELYON16                      & \begin{tabular}[c]{@{}l@{}}399 WSIs - 270/129 in train/test set\\ Train - 70/100 positive/negative WSIs\\ Test - 49/80 positive/negative WSIs\end{tabular}                                               & {\begin{tabular}[c]{@{}c@{}}Application of HistoROI\\ as WSI pre-processing\end{tabular}} \\ \cline{1-2}
\multicolumn{1}{|l|}{TCGA-Lung} & \begin{tabular}[c]{@{}l@{}}1034 WSIs - 634/404 in train/test set\\ Train - 316/318 adeno/squamous WSIs\\ Test - 211/193 adeno/squamous WSIs\end{tabular}                                                 &                                                                                                           \\ \hline
\end{tabular}
\caption{Summary of the datasets used in this study}
\label{datasets}
\end{table*}

The datasets used in this study are summarised in Table ~\ref{datasets}. BRIGHT is a dataset with 503 WSIs labeled as cancerous, pre-cancerous, or non-cancerous~\cite{bright}. We have used the patches from 50 WSIs of the BRIGHT dataset for the training of HistoROI. These WSIs were selected carefully to capture variations for better generalization. The detailed process of selecting these WSIs and patch-level data preparation is described in Section~\ref{sec_methods}. This dataset contains more than 2 million labeled patches. Each patch is assigned one of the six possible labels -- epithelium, stroma, lymphocytes, adipose, artifacts, or miscellaneous. WSIs from all the classes in the BRIGHT dataset were considered for the preparation of patch-level datasets. The artifact class includes patches with out-of-focus areas, tissue folds, cover slips, air bubbles, pen markers, and extreme overstaining or understaining. 

CRC dataset~\cite{crc100k} is a colon cancer dataset of 100,000 patches of size 224x224. The patches in this dataset are segregated into nine different classes, viz. normal (NORM), tumor (TUM), mucin (MUC), muscle (MUS), background (BACK), debris (DEB), adipose (ADI), lymphocytes (LYM) and stroma (STR). Class-wise distribution of patches in this dataset can be found in Table~\ref{datasets}. 

Classification models trained using weakly supervised learning are trained and validated on CAMELYON16 and TCGA-Lung datasets. CAMELYON16 is a dataset of 399 WSIs of sentinel lymph nodes associated with breasts with 270 images for training and validation and 129 WSIs for testing. The TCGA-Lung dataset contains 1034 WSIs. Further details of these datasets are shown in Table~\ref{datasets}. 

To validate the effectiveness of HistoROI for identifying artifacts in WSIs, we manually annotated a dataset of 93 WSIs to delineate the foreground tissue regions (excluding artifacts). These WSIs were selected from four different organs and contain various artifacts and a wide range of stain intensities and colors. All the WSIs were downloaded from the TCGA data portal. This dataset contains 27 WSIs from breast tissue, 21 WSIs from lung tissue, and 21 and 24 WSIs from kidney and prostate tissues, respectively. This test dataset was selected from a different source and contains three additional organs to test if the model can strongly generalize after being trained solely on WSIs from the BRIGHT breast tissue dataset~\cite{bright}.

\section{METHODOLOGY}\label{sec_methods}

\begin{figure*}[h]
	\centering
		\includegraphics[width=\textwidth]{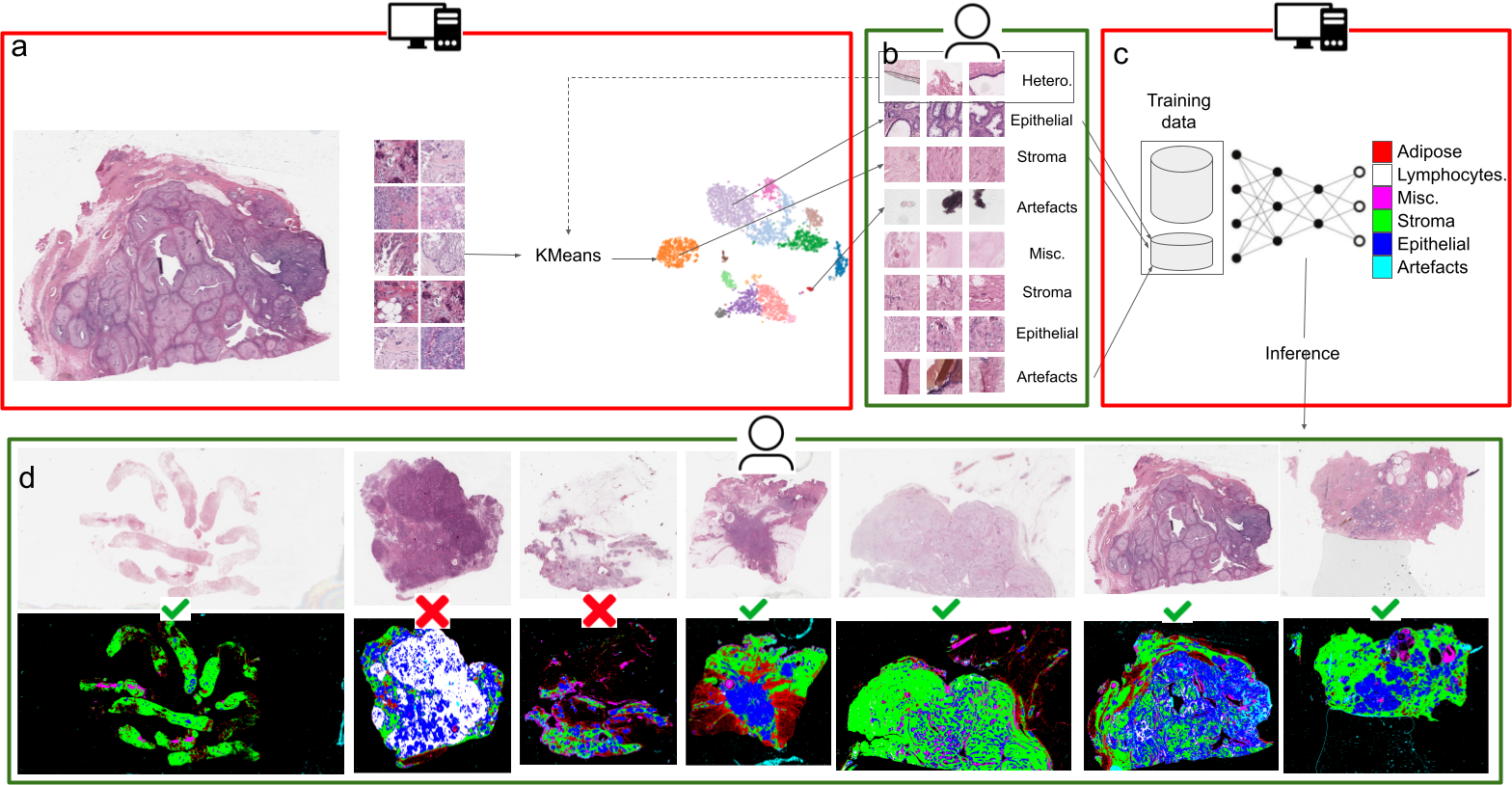}
	\caption{Human-in-the-loop training pipeline for HistoROI: Actions in red boxes are automatic while those in green boxes are manual. (a) Embeddings of the patches of WSI are divided into clusters. (b) Clusters are manually annotated. Heterogeneous clusters are re-clustered (shown using dotted line) (c) Newly annotated data is added to previously annotated data and HistoROI is trained with the updated data. (d) The trained model is applied (tested) on multiple WSIs. WSIs with poor performance are manually identified (denoted by X) and annotated for the next iteration of training.}
	\label{train-pipeline}
\end{figure*}

HistoROI is a ResNet18-based six-class classifier that assigns a pathology-relevant label to a patch extracted from a WSI. To reduce annotation time for a WSI, we have taken advantage of features from shallow layers of pre-trained CNNs. We have leveraged a novel human-in-loop active learning approach to ensure diversity in the annotated datasets. With the help of k-means clustering and easy-to-use user interfaces, the annotation of each WSI (around 40,000 patches) takes less than 15 minutes. The details of dataset preparation and training of HistoROI are explained in this section.

\textbf{Annotation of a single WSI}: 
We have used a clustering-based approach to annotate patches of WSIs. We extract patches of size 256x256 at 10x magnification (approx. $1\mu m \times 1\mu m$ per pixel dimension) from a WSI. Patches with mostly white pixels are not considered for data annotation by removing those with more than 95\% of pixels with average RGB values greater than 230. The rest of the patches are given as an input to an EfficientNet-b0~\cite{efficientnet} CNN architecture pre-trained on ImageNet to extract instance (patch) features. Specifically, we extract a tensor with 40 feature channels from block-2 of EfficientNet-b0 and use global average pooling to get a 40-dimensional feature vector for a patch. We have tried extracting features from other layers of EfficientNet-b0 and different versions of EfficientNets as well as ResNets and observed that clusters created using features of block-2 of EfficientNet-b0 are semantically more homogeneous. While segregating patches of WSI, we mostly care about local features, such as texture, to distinguish between different types of tissue regions, such as stroma or nucleus-dense epithelium. The diameter of nuclei at a 10x magnification level is less than a few pixels, which can be learned by such local features. Therefore, we hypothesize that the identification of classes of our interest can better be done with local features extracted by block-2 of EfficientNet-b0 than the global features we get from deeper layers of pre-trained CNNs. 

Features of all the patches in a WSI are then clustered using a k-means clustering algorithm, for which fast implementation is available in scikit-learn \cite{sklearn_api}. We learn 32 clusters and sample 25 patches from a cluster in the form of a 5x5 grid for annotation. Upon manual inspection if a grid of 5x5 patches turns out to be homogeneous -- that is, it contains patches of the same class -- then we assign one of the following labels to all the patches corresponding to that cluster: 1. Epithelium 2. Stroma 3. Lymphocytes 4. Adipose 5. Artifact, and 6. Miscellaneous. Patches from heterogeneous clusters are pooled together and re-clustered into 32 new clusters using another round of k-means, and the labeling process is repeated, as shown by a dotted line in Figure~\ref{train-pipeline}. Patches belonging to heterogeneous clusters after the second stage of this human-in-the-loop labeling are discarded. Generally, less than 5\% of the patches are discarded from a WSI after the second round. 

We initially created a training dataset with 20 WSIs and grew this dataset as the training process progressed. The annotation of each WSI took around 15 minutes.

\textbf{Diversity-based expansion of annotations}:
WSI annotation is a time-consuming task. Therefore it is important to annotate a diverse set of WSIs instead of wasting time and effort annotating multiple WSIs with similar visual features. Diversity and label uncertainty are two major guiding principles of active learning approaches~\cite{sener2017active}. We trained the model and annotated additional WSIs with an approach inspired from active learning to address this problem. 

We initially trained a ResNet18-based~\cite{resnet} six-class classifier with the data from 20 WSIs. We used patches from 15 WSIs for training and 5 WSIs for validation. The classifier was trained on NVIDIA-3090Ti with a batch size of 128. Cross entropy loss was optimized with Adam optimiser~\cite{adam}, and the model with the least validation loss was applied on all the WSIs from the BRIGHT dataset. QuPath-compatible~\cite{qupath} visualisations were created for all images. WSIs and their predicted patch classifications were visually analyzed using QuPath. WSIs with poor performance indicate that the current WSI is out-of-distribution with respect to the training dataset. That is, such WSIs has high label uncertainty, and it should be a good candidate for annotation in the next round to maximize information gain while minimizing annotation effort. These WSIs were manually identified, annotated and added to the training dataset for further fine-tuning of the classifier. We repeated this cycle three times, adding 10 WSIs to the training dataset at a time. Thus, we created a training dataset with 50 carefully selected and annotated WSIs containing enough variation for generalization. The whole process of human-in-the-loop training is summarised in Figure~\ref{train-pipeline}.

\textbf{Classification models}: 
As a representative of the downstream models that can benefit from HistoROI, we trained widely used weakly supervised learning algorithm called CLAM for histopathology~\cite{clam}. CLAM is trained with a multiple-instance learning paradigm, where the feature of a patch is used as an instance, and the set of these instances from a WSI (either all patches or a subset thereof) is used to create a bag. CLAM has shown improved performance on multiple publicly available datasets~\cite{camelyon, tcga}. We trained and validated CLAM in multiple patch pre-filtering settings to verify the effectiveness of the proposed classification method and model in improving the effectiveness of weakly supervised learning.

\section{RESULTS AND DISCUSSION}
\label{sec_results}
\begin{figure*}[h]
	\centering
		\includegraphics[width=\textwidth]{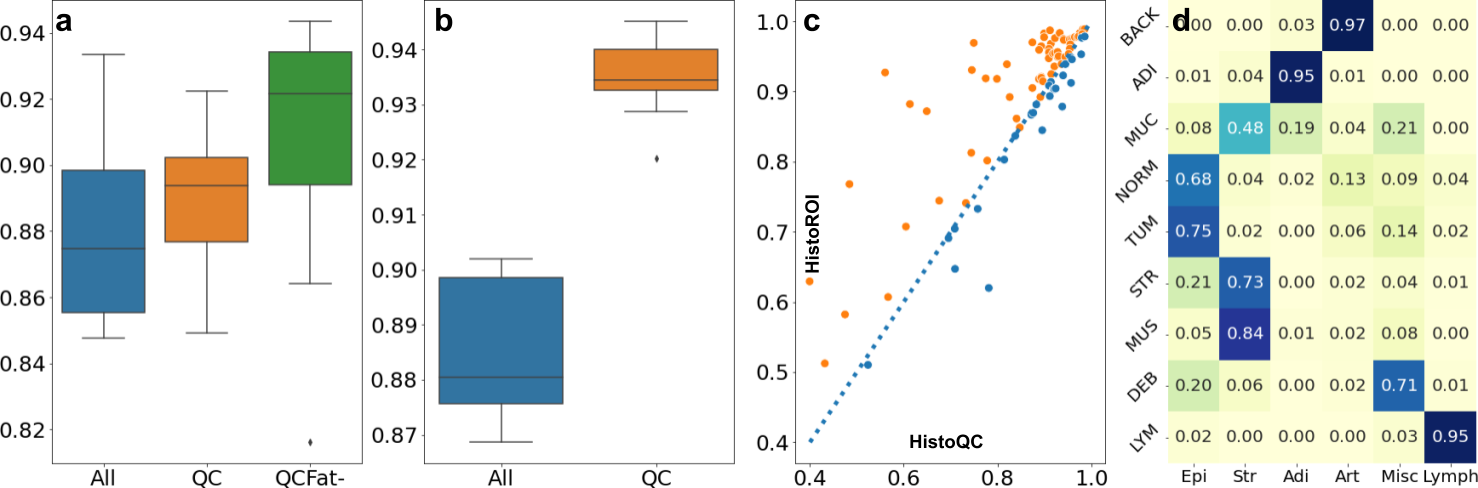}
	\caption{\textbf{HistoROI quantitative results:} (a) Box plot of AUCs on the test split of the CAMELYON16 dataset shows improvement in performance when WSIs are pre-processed with HistoROI. (b) Box plot of AUCs for the TCGA-LUNSC dataset. (c) Comparison of HistoROI with HistoQC~\cite{histoqc} for QC using TCGA-4Org dataset shows that HistoROI's Dice score is better than HistoQC for 65 WSIs out of 93 WSIs. (d) HistoROI predicts 77\% of patches correctly in the CRC-100k without being trained on this dataset.}
	\label{quantitative_results}
\end{figure*}
\subsection{HistoROI as WSI pre-processing tool}
We first show the applicability of HistoROI as a pre-processing tool for a weakly supervised learning algorithm -- CLAM~\cite{clam}. Even though weakly supervised learning methods are supposed to filter out irrelevant regions, including artifacts, automatically, we show that these methods can still benefit from data pre-filtering based on HistoROI. As a baseline, we train CLAM with the bag prepared by features of all the patches from a WSI. Then, for comparison, we filter out patches from the bags based on the prediction of HistoROI and analyze changes in the performance of CLAM applied to this filtered data. More specifically, we perform two experiments. The first experiment shows the utility of HistoROI at train and test times. In this experiment, we use the CAMELYON dataset to demonstrate improvement in the performance of CLAM when data filtering is used at both training and test times. In the second experiment, we try to emulate a real-world scenario. Generally, training data goes through a few iterations of labeling/annotations before being fed to the deep learning algorithm. Through this process, training data is cleaned and curated hence expected to contain lesser anomalies. On the other hand, test data can not be manually curated. Hence, in the second experiment, we do not use data filtering on training data but filter the patches in test WSI using HistoROI at inference. We have used the TCGA-Lung dataset to distinguish between adenocarcinoma and squamous cell carcinoma for this experiment and have shown significant improvement in the performance when HistoROI is used for data filtering. This experiment also shows the possibilities of using HistoROI on algorithms that are independently developed. We have created bags for CLAM in three different ways in these experiments. 1. Bags with the features from all the non-white patches (“All”), 2. Bags without patches identified as artefacts by HistoROI (“QC”), and 3. Bags without artefacts and adipose patches (“QCFat-”).

\begin{figure}[h]
	\centering
		\includegraphics[width=0.9\textwidth]{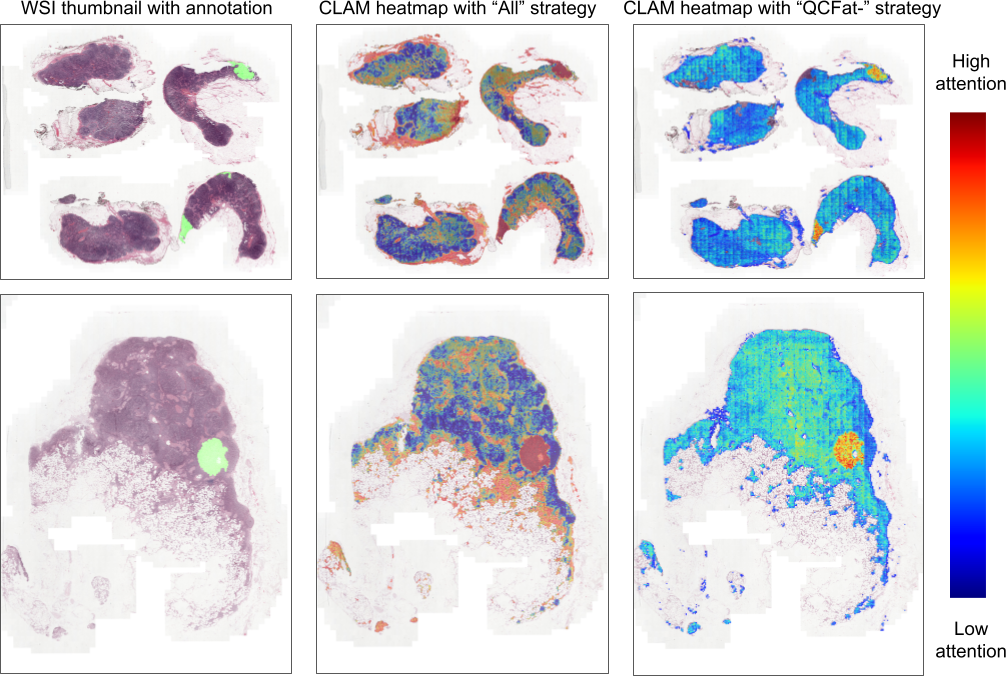}
	\caption{\textbf{Qualitative results on CAMELYON16:} CLAM~\cite{clam} trained on regions filtered by HistoROI generates better attention maps. Metastasised region is annotated with cyan color in the first column. CLAM attention map for the "All" strategy is scattered compared to the accurate and specific one generated using the "QCFat--" strategy.}
	\label{qc-camelyon}
\end{figure}

We trained CLAM ten times with different train-validation splits on the CAMELYON16 dataset. We used the same test split created by the CAMELYON16 challenge organizers for testing. We used 80\% of training WSIs for training and the remaining 20\% for validation to tune hyper-parameters. The mean AUC on test data was observed to increase from 0.88 to 0.90, while the median AUC increased from 0.87 to 0.92 as shown in Figure~\ref{quantitative_results} (a). Filtering out the patches with artifacts clearly improves the performance compared to using all the patches. Further removing patches with adipose tissue increased performance even further. We tried various other patch selection criteria along with different color normalization methods~\cite{vahadane, abhijeet} and did not notice a significant change in test accuracy. Interestingly, we noticed a significant reduction in test accuracy when we used only epithelial patches. Upon closer inspection, we noticed that for a few WSIs, most patches contain both epithelial nuclei and lymphocytes. HistoROI predicted those patches as lymphocytes, and hence informative patches were filtered out from the training algorithm. This observation highlights that assigning multiple labels for a patch can be more useful than assigning a single one. 

For our next experiment, we assessed the impact of using HistoROI for a scenario in which the training data is relatively cleaner than the test data, which is to be expected in practice. We hypothesize that if real-world data is filtered using a reliable quality control pipeline, the performance of the model trained on relatively cleaner data can be improved during test time. To demonstrate this, we have conducted experiments with TCGA-Lung cancer data. For this experiment, we intentionally prepared a test split with WSIs that contain a relatively higher proportion of artifacts. The list of WSIs in the test split is listed on the webpage with the code of our method \href{https://github.com/abhijeetptl5/historoi}{HistoROI}. We trained CLAM ten times with different train-validation splits and analysed the change in performance at test time with and without data filtering. Quantitative results are shown as boxplots in Figure~\ref{quantitative_results} (b). We observe that the performance of CLAM significantly improves when inferred WSI is pre-processed using HistoROI.

Along with better quantitative performance, the model trained with the assistance of HistoROI produces better explainability. We have visualized attention maps for a few WSIs trained with the “All” strategy and “QCFat--” strategies for the CAMLEYON dataset. As shown in Figure~\ref{qc-camelyon}, the attention map created by CLAM when trained with filtered patches is in concordance with the annotations, whereas CLAM trained with all the patches gives high attention to arbitrary locations, including regions containing adipose. According to our domain knowledge, features of the adipose region do not correlate with the presence of metastasis in lymph nodes. But high attention on the adipose region indicates the potential data bias with the correlation between the adipose region and the presence of metastasis. We have removed this potential bias by filtering out adipose region from analysis and observed better results with the ``QCFat--" strategy.

\subsection{Validation of HistoROI on CRC-100k dataset}

\begin{figure}[h]
	\centering
		\includegraphics[width=0.9\textwidth]{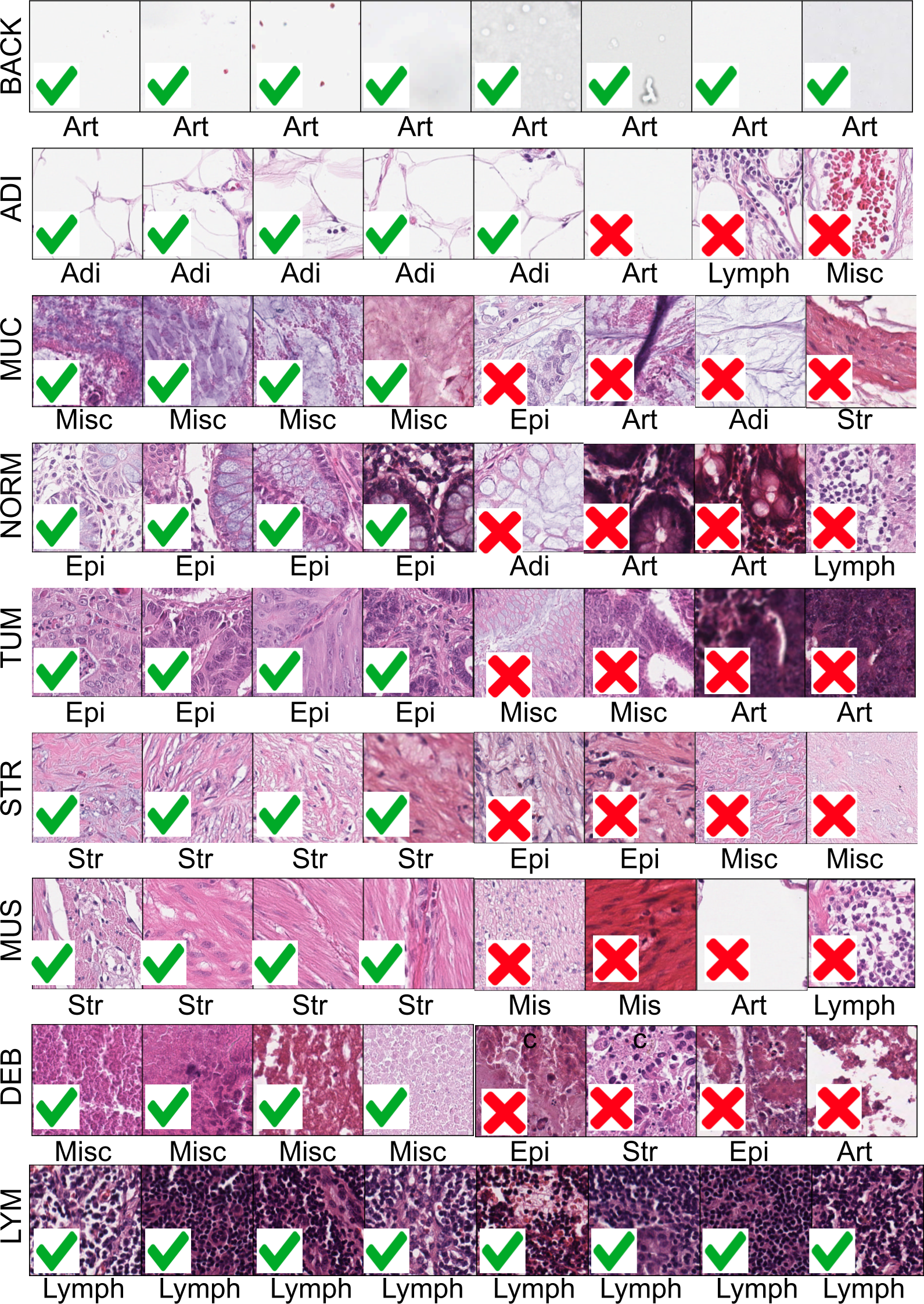}
	\caption{\textbf{Qualitative results on CRC-100k\cite{crc100k}:} Example predictions (column labels) on CRC-100k dataset by HistoROI for various classes (row labels) shows that misclassified patches (with red crosses) tend to have background, staining issues, or multiple classes, which are not present in the correct classified patches (with green ticks). Here BACK stands for the background, ADI for adipose, MUC for mucus, NORM for normal, TUM for tumor, STR for stroma, MUS for muscle, DEB for debris, and LYM for lymphocytes.}
	\label{crc-results}
\end{figure}

We applied HistoROI to the CRC dataset to validate its generalization on the colon cancer dataset even though it was trained on the BRIGHT breast cancer dataset. The CRC dataset contains a few classes exclusive to the prediction classes of HistoROI. We expect the adipose (ADI) class of the CRC dataset to be predicted as adipose, normal (NORM) as epithelium, background (BACK) as artifacts, lymphocytes (LYM) as lymphocytes, tumor (TUM) as epithelium, mucus (MUC) and debris (DEB) as miscellaneous, and stroma (STR) and muscle (MUS) as stroma. HistoROI predicted 77\% of images in the CRC-100k dataset correctly. HistoROI predicted more than 90\% of patches in LYM, ADI, and BACK classes correctly. More than 70\% of patches from the DEB class were correctly classified as miscellaneous. Twenty percent of DEB patches were predicted as epithelium.
Most of these wrongly predicted patches contain a few epithelial nuclei in debris. For normal class (NORM), 13\% of patches were predicted as artifacts. Our analysis indicates that few of these wrongly predicted patches contain blurry regions, and the remaining patches are overstained and thus contain some artifacts. Similarly, tumor patches misclassified as miscellaneous were observed to contain debris. Most classification errors were observed in the mucus (MUC) class. Most of the mucus patches were predicted as stroma by HistoROI as the two classes share visual features. Additionally, HistoROI has not seen patches with mucus while training on the BRIGHT dataset. Finally, it can be hard to distinguish between stroma and mucus in a small patch to begin with. The dataset contains patches of size 224x244 at 20x, corresponding to 112x112 when downscaled to 10x for HistoROI. 

Figure~\ref{quantitative_results} (d) shows quantitative results and Figure~\ref{crc-results} highlights a few correctly and wrongly predicted patches for the CRC dataset. Analysis of these results can lead to better design of models and datasets. For example, a few patches in the NORM class that were incorrectly predicted as artifacts contain artifacts in the form of blur or dark stains. These observations highlight subjectivity in labeling histopathology patches. We propose that rather than assigning hard labels to a patch, a better strategy would be to assign multiple labels or assign a level of a particular class along with the label. Since we have trained a HistoROI with patches from breast cancer dataset, the low accuracy for the MUC class shows that patches from multiple diverse organs should be used to train a robust model. The human-in-the-loop paradigm presented here provides an easy way to expand the dataset and the model generalization.

\subsection{HistoROI as QC tool}

\begin{figure}[h]
	\centering
		\includegraphics[width=0.9\textwidth]{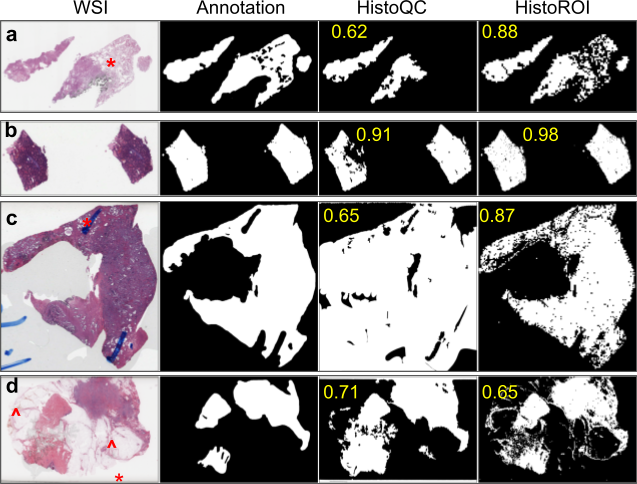}
	\caption{\textbf{Qualitative results on TCGA-4Org dataset:} Foreground detection (with Dice score) for a few examples WSIs show improved foreground detected by HistoROI over HistoQC.}
	\label{qc-results}
\end{figure}

The background (BACK) class in CRC dataset contains mostly white patches, which does not effectively validate predictions for the artifact class of HistoROI. To address this issue, we annotated the foreground region in 93 WSIs and compared predictions made by HistoROI with it. Annotated dataset contains WSIs from four different organs (27 from breast, 21 from lung, 21 from kidney, and 27 from prostate). We also compared predictions made by HistoQC~\cite{histoqc} -- a popular histopathology quality control tool -- with HistoROI. The mean Dice score over WSIs between HistoROI and hand annotations was 0.87, whereas, for HistoQC, it was 0.83. A few qualitative results are shown in Figure~\ref{qc-results}. The performance of HistoROI was better on 65 WSIs out of 93 WSIs, as shown in Figure~\ref{quantitative_results} (c).

According to our observations, HistoQC tends to identify the region with relatively less dense tissue as fat . Also, HistoQC fails to distinguish between foreground and background when background pixels are greyish. Because of this, the performance of HistoQC degrades in the presence of air bubbles (Figure~\ref{qc-results} (c)) and coverslip-related artifacts ). HistoROI performs better compared to HistoQC for the scenarios mentioned above. 

On the other hand, HistoQC detects pen marks better than the proposed HistoROI. Further analysis of training data for HistoROI showed that it contains only one WSI with pen marks. These pen marks were also observed to be outside the tissue region. Hence, in Figure~\ref{qc-results} (c) , pen marks outside the tissue region are correctly identified as background by HistoROI. In contrast, the pen marker inside the tissue region is not delineated properly by HistoROI. Also, for a few WSIs, the Dice score for HistoQC was observed to be greater than that for HistoROI, even if visually, HistoROI performance is better. This uncovers difficulties in annotating WSIs. One such example is shown in Figure~\ref{qc-results} (d). Detailed analysis of the prediction of HistoROI can lead to guidelines for building the next generation of a QC solution, which we think will simply involve sampling from more diverse training data, such as one that contains more organs and slides with marker pen artifacts. Though the predictions of HistoROI are pixelated (because its model is based on patches), the overall Dice score is better than HistoQC. This indicates that the patch-based dataset contains enough variation for recognizing most artifacts. A dataset for semantic segmentation with current data can be useful for designing the next generation of QC solutions for histopathology images.

\section{Conclusion} \label{sec_conclusion}

In this study, we have developed a human-in-the-loop active learning paradigm to prepare a patch-level dataset for classifying pathology patches. With the prepared data, we have trained a ResNet18-based six-class classifier called HistoROI. To validate the generalization of HistoROI, we tested it for classifying colon tissue patches using the CRC dataset. We also explored the potential of HistoROI as a quality control (QC) tool for WSIs. Further, we investigated the use case for HistoROI for pre-processing WSIs in classification pipelines based on the widely-used weakly supervised learning classification algorithm –- CLAM. Through our experiments, we have shown improvement in the performance of CLAM when HistoROI is used for pre-filtering instances (patches) for each bag (WSI). We have also shown improvements over the widely used quality control tool, HistoQC, for the artifact detection. HistoROI was trained using the BRIGHT dataset, which is a breast cancer dataset. We conducted validation experiments utilizing datasets from multiple data centers and organs, such as TCGA, CAMELYON and CRC-100k. The results demonstrate that HistoROI can generalize effectively on high resolution whole slide images. Nonetheless, it remains to be determined whether HistoROI can perform equally well on manually acquired microscope images, which may exhibit significant domain shifts.

Our analysis of HistoROI has brought to light certain limitations that impede its ability to accurately classify patches of histopathology images containing multiple tissue segments. These limitations stem from the single label assignment strategy to annotate the dataset, which restricts HistoROI to assigning a single label to each patch of a WSI. To address this issue, we propose several solutions, including optimizing patch size for annotation, employing segmentation masks instead of patch-wise labeling, and providing pathologists with easy-to-use interactive segmentation and active learning tools for faster annotation. Moreover, we suggest asking annotators to identify the presence of a certain class in a given patch rather than selecting from a set of labels, as well as utilizing multiple annotators to improve accuracy of annotations. By implementing these measures, we believe that HistoROI can be improved and better utilized in the field of histopathology analysis. Also, we propose that the performance of HistoROI can be improved by training it on multiple organs to develop more robust and generalizable models. For the artifact detection task, we have noticed that the performance of HistoROI does not detect pen marks well due to the lack thereof in its training data. Further analysis should be carried out to investigate cases where the proposed model can not predict artifacts correctly. This can lead to the preparation of better datasets using which the next version of HistoROI can be trained and released.

\section*{Acknowledgment}
This project has received funding from the Department of Biotechnology Ministry of Science and Technology, Government of India under grant agreement No. BT/PR32348/AI/133/25/2020 (Project name: Imaging Biobank for cancer).


\printbibliography

@INPROCEEDINGS{breakhis,
  author={Spanhol, Fabio Alexandre and Oliveira, Luiz S. and Petitjean, Caroline and Heutte, Laurent},
  booktitle={2016 International Joint Conference on Neural Networks (IJCNN)}, 
  title={Breast cancer histopathological image classification using Convolutional Neural Networks}, 
  year={2016},
  volume={},
  number={},
  pages={2560-2567},
  doi={10.1109/IJCNN.2016.7727519}}

@misc{bach, 
title={BACH Dataset: Grand Challenge on Breast Cancer Histology images}, 
volume={56}, 
DOI={10.5281/zenodo.3632035}, 
journal={Medical Image Analysis}, 
publisher={Zenodo}, 
author={Polónia, António and Eloy, Catarina and Aguiar, Paulo}, 
year={2019}, 
month={May}, 
pages={122–139} }

@misc{clam,
  doi = {10.48550/ARXIV.2004.09666},
  url = {https://arxiv.org/abs/2004.09666},
  author = {Lu, Ming Y. and Williamson, Drew F. K. and Chen, Tiffany Y. and Chen, Richard J. and Barbieri, Matteo and Mahmood, Faisal},
  title = {Data Efficient and Weakly Supervised Computational Pathology on Whole Slide Images},
  publisher = {arXiv},
  year = {2020}, 
  copyright = {arXiv.org perpetual, non-exclusive license}
}

@article{bright,
  author={Brancati, Nadia and De Pietro, Giuseppe and Riccio, Daniel and Frucci, Maria},
  journal={IEEE Access}, 
  title={Gigapixel Histopathological Image Analysis Using Attention-Based Neural Networks}, 
  year={2021},
  volume={9},
  number={},
  pages={87552-87562},
  doi={10.1109/ACCESS.2021.3086892}}

@article{sener2017active,
  title={Active learning for convolutional neural networks: A core-set approach},
  author={Sener, Ozan and Savarese, Silvio},
  journal={arXiv preprint arXiv:1708.00489},
  year={2017}
}

@article{selfsupwsi,
  author={Chen, Chengkuan and Lu, Ming Y. and Williamson, Drew F. K. and Chen, Tiffany Y. and Schaumberg, Andrew J. and Mahmood, Faisal},
  journal={Nature Biomedical Engineering}, 
  title={Fast and scalable search of whole-slide images via self-supervised deep learning}, 
  year={2022},
  doi={10.1038/s41551-022-00929-8}}

@article{tcga,
  author={Chang, Kyle et. al.},
  journal={Nature Genetics}, 
  title={The Cancer Genome Atlas Pan-Cancer analysis project}, 
  year={2013},
  doi={10.1038/ng.2764}}

@article{stroma,
  author={Xu, Sj. and Zhang, Sy. and Dong, Ly. et al.},
  journal={BMC Cancer}, 
  title={Dynamic survival analysis of gastrointestinal stromal tumors (GISTs): a 10-year follow-up based on conditional survival.}, 
  year={2021},
  volume={21},
  doi={10.1186/s12885-021-08828-y}}

@article{stroma2,
  author={Millar EK and Browne LH and Beretov J and Lee K and Lynch J and Swarbrick A and Graham PH.},
  journal={Cancers (Basel)}, 
  title={Tumour Stroma Ratio Assessment Using Digital Image Analysis Predicts Survival in Triple Negative and Luminal Breast Cancer.}, 
  year={2020},
  volume={12},
  doi={10.3390/cancers12123749}}

@INPROCEEDINGS{tiger,
  author = {Mart van Rijthoven},
  title = {tumor-infiltrating lymphocytes (TILs) in H\&E breast cancer},
  year = 2021,
  url = {https://tiger.grand-challenge.org/Home/},
  urldate = {2022-09-09}
}

@online{grandc,
  title = {Grand Challenges},
  url = {https://grand-challenge.org/},
  urldate = {2022-09-09}
}

@misc{adam,
  doi = {10.48550/ARXIV.1412.6980},
  url = {https://arxiv.org/abs/1412.6980},
  author = {Kingma, Diederik P. and Ba, Jimmy},
  title = {Adam: A Method for Stochastic Optimization},
  publisher = {arXiv},
  year = {2014},
}

@ARTICLE{jbhi,
  author={Wright, Alexander I. and Dunn, Catriona M. and Hale, Michael and Hutchins, Gordon G. A. and Treanor, Darren E.},
  journal={IEEE Journal of Biomedical and Health Informatics}, 
  title={The Effect of Quality Control on Accuracy of Digital Pathology Image Analysis}, 
  year={2021},
  volume={25},
  number={2},
  pages={307-314},
  doi={10.1109/JBHI.2020.3046094}}

@ARTICLE{stresstest,
  author={Schömig-Markiefka, B. and Pryalukhin, A. and Hulla, W. et al.},
  journal={Modern Pathology}, 
  title={Quality control stress test for deep learning-based diagnostic model in digital pathology.}, 
  year={2021},
  volume={34},
  doi={10.1038/s41379-021-00859-x}}

@INPROCEEDINGS{abhijeet,
  author={Patil, Abhijeet and Talha, Mohd. and Bhatia, Aniket and Kurian, Nikhil Cherian and Mangale, Sammed and Patel, Sunil and Sethi, Amit},
  booktitle={2021 IEEE 18th International Symposium on Biomedical Imaging (ISBI)}, 
  title={Fast, Self Supervised, Fully Convolutional Color Normalization Of H\&E Stained Images}, 
  year={2021},
  volume={},
  number={},
  pages={1563-1567},
  doi={10.1109/ISBI48211.2021.9434121}}

@ARTICLE{vahadane,
  author={Vahadane, Abhishek and Peng, Tingying and Sethi, Amit and Albarqouni, Shadi and Wang, Lichao and Baust, Maximilian and Steiger, Katja and Schlitter, Anna Melissa and Esposito, Irene and Navab, Nassir},
  journal={IEEE Transactions on Medical Imaging}, 
  title={Structure-Preserving Color Normalization and Sparse Stain Separation for Histological Images}, 
  year={2016},
  volume={35},
  number={8},
  pages={1962-1971},
  doi={10.1109/TMI.2016.2529665}}

@ARTICLE{histoqc,
  author={Janowczyk A and Zuo R and Gilmore H and Feldman M and Madabhushi A.},
  journal={JCO Clin Cancer Inform.}, 
  title={HistoQC: An Open-Source Quality Control Tool for Digital Pathology Slides.}, 
  year={2019}}

@ARTICLE{histoqc2,
  author={Chen Y and Zee J and Smith A and Jayapandian C and Madabhushi A and Barisoni L and Janowczyk A.},
  journal={J Pathol.}, 
  title={Assessment of a computerized quantitative quality control tool for whole slide images of kidney biopsies.}, 
  year={2021}}

@ARTICLE{pathprofiler,
  author={Haghighat, M. and Browning, L. and Sirinukunwattana, K. et al. },
  journal={Sci Rep}, 
  title={Automated quality assessment of large digitised histology cohorts by artificial intelligence.}, 
  year={2022},
  doi={10.1038/s41598-022-08351-5}}

@misc{crc100k, 
title={100,000 histological images of human colorectal cancer and healthy tissue},
DOI={10.5281/zenodo.1214456}, 
publisher={Zenodo}, 
author={Kather, Jakob Nikolas and Halama, Niels and Marx, Alexander}, 
year={2018}, month={Apr}}

@ARTICLE{camelyon,
  author={Bándi Péter and Geessink Oscar and Manson Quirine and Van Dijk and van der Laak, Jeroen and Litjens Geert},
  journal={IEEE Transactions on Medical Imaging}, 
  title={From Detection of Individual Metastases to Classification of Lymph Node Status at the Patient Level: The CAMELYON17 Challenge}, 
  year={2019},
  volume={38},
  number={2},
  pages={550-560},
  doi={10.1109/TMI.2018.2867350}}

@article{resnet,
  author    = {Kaiming He and
               Xiangyu Zhang and
               Shaoqing Ren and
               Jian Sun},
  title     = {Deep Residual Learning for Image Recognition},
  journal   = {CoRR},
  volume    = {abs/1512.03385},
  year      = {2015},
  url       = {http://arxiv.org/abs/1512.03385},
  eprinttype = {arXiv},
  eprint    = {1512.03385},
  bibsource = {dblp computer science bibliography, https://dblp.org}
}

@article{efficientnet,
  doi = {10.48550/ARXIV.1905.11946},
  url = {https://arxiv.org/abs/1905.11946},
  author = {Tan, Mingxing and Le, Quoc V.},
  title = {EfficientNet: Rethinking Model Scaling for Convolutional Neural Networks}, 
  publisher = {arXiv},
  year = {2019},
}

@ARTICLE{qupath,
  author={Bankhead P and Loughrey MB  and Fernández JA and Dombrowski Y and Salto-Tellez M and Hamilton PW},
  journal={Sci Rep.}, 
  title={QuPath: Open source software for digital pathology image analysis.}, 
  year={2017},
  doi={10.1038/s41598-017-17204-5}}

@article{jco_nikhil,
author = {Kanse, Abhiraj S. and Kurian, Nikhil C. and Aswani, Himanshu P. and Khan, Zakia and Gann, Peter H. and Rane, Swapnil and Sethi, Amit},
title = {Cautious Artificial Intelligence Improves Outcomes and Trust by Flagging Outlier Cases},
journal = {JCO Clinical Cancer Informatics},
volume = {},
number = {6},
pages = {e2200067},
year = {2022},
doi = {10.1200/CCI.22.00067},
}

@article{jpath_deepak,
author = {Deepak Anand and Kumar Yashashwi and Neeraj Kumar and Swapnil Rane and Peter H Gann and Amit Sethi},
title = {Weakly supervised learning on unannotated H\&E-stained slides predicts BRAF mutation in thyroid cancer with high accuracy},
journal = {The Journal of Pathology},
year = {2021},
doi = {10.1002/path.5773},
}

@inproceedings{sklearn_api,
  author    = {Lars Buitinck and Gilles Louppe and Mathieu Blondel and
               Fabian Pedregosa and Andreas Mueller and Olivier Grisel and
               Vlad Niculae and Peter Prettenhofer and Alexandre Gramfort
               and Jaques Grobler and Robert Layton and Jake VanderPlas and
               Arnaud Joly and Brian Holt and Ga{\"{e}}l Varoquaux},
  title     = {{API} design for machine learning software: experiences from the scikit-learn
               project},
  booktitle = {ECML PKDD Workshop: Languages for Data Mining and Machine Learning},
  year      = {2013},
  pages = {108--122},
}
\end{document}